\documentclass[twocolumn,showpacs,prb,amsmath,amssymb]{revtex4}
\usepackage{graphicx}
\usepackage{amssymb}

\begin{document}
\title{Vibrational properties of inclusion complexes: the case of
  indomethacin-cyclodextrin}
\author{Barbara Rossi, Paolo Verrocchio and Gabriele Viliani}
\affiliation{Dipartimento di Fisica, Universit\`a di Trento, 38050
Povo, Trento, Italy,\\ and\\ INFM CRS-SOFT, c/o Universit\`a di Roma
"La Sapienza", 00185, Roma, Italy.}
\author{Giorgina Scarduelli, Graziano Guella and Ines Mancini}
\affiliation{Laboratorio di Chimica Bioorganica, Dipartimento di
Fisica, Universit\`a di Trento,\\ 38050 Povo, Trento, Italy.}

\date{\today}
\begin{abstract}
  Vibrational properties of inclusion complexes with cyclodextrins
  are studied by means of Raman spectroscopy and numerical simulation.
  In particular, Raman spectra of the non-steroidal, anti-inflammatory drug
  indomethacin undergo notable changes in the energy range between
  1600 and 1700 cm$^{-1}$ when inclusion complexes with cyclodextrins are formed.
  By using both \emph{ab initio} quantum chemical calculations and molecular
  dynamics, we studied how to relate such changes to the geometry
  of the inclusion process,
   disentangling single-molecule effects, from changes in the solid state structure or
  dimerization processes.
\end{abstract}
 \maketitle

\section{Introduction}
Inclusion complexes with cyclodextrins have been little studied by
means of Raman spectroscopy, although the usefulness of this
technique in such systems is now acknowledged. \cite{b.san,
b.fini} Cyclodextrins (CD) are a family of natural or
synthetically modified cyclic molecules, consisting typically of
six ($\alpha$-CD), seven ($\beta$-CD) or eight ($\gamma$-CD)
glucopyranose units. In water, they take on the peculiar
3-dimensional structure of a truncated cone, with a slightly
soluble outer surface and a hydrophobic central cavity, whose size
depends on the number of glucose units of the molecule. A
remarkable property of CD in aqueous solution, is their ability to
form host-guest inclusion complexes with a wide variety of organic
and inorganic molecules, provided that the guest molecule is less
polar than water.\cite{b.li} In particular, inclusion complexes of
CD with non-polar drugs are a topic of current interest, because
these non-covalent complexes increase not only the aqueous
solubility of drugs, but also their chemical stability and
bioavailability.

Such complexes have been usually investigated in aqueous solution
by ultra-violet and visible absorption spectroscopy, fluorescence
techniques, and NMR spectroscopy, while X-ray and neutron
diffraction are typically utilized in the solid state, but the
latter techniques (unlike Raman) require crystalline samples. The
effect of the inclusion process on the guest molecules has also
been the subject of some Raman studies;\cite{b.raman1, b.raman2,
b.raman3} in particular, Raman and FT-IR investigations of the
interactions between CD and some non-steroidal anti-inflammatory
agents (like diclofenac sodium, \cite{b.iliescu} ketoprofen
\cite{b.choi} and amorphous piroxicam \cite{b.redenti}), have
shown significant changes in the vibrational spectrum of the
complexed guest molecules with respect to the free ones.

\begin{figure}
\centering \includegraphics[width=.8\columnwidth]{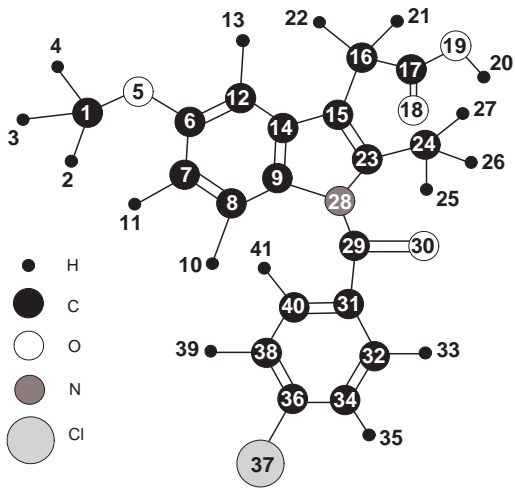}
\caption{Chemical structure of indomethacin, with numbering of
atoms adopted for the molecular dynamics computations.}
 \label{f.1}
\end{figure}

Among the non-steroidal anti-inflammatory drugs, indomethacin
(IMC, Fig.~\ref{f.1}) is widely used as an analgesic drug in the
treatment of rheumatoid arthritis, as well as in other
degenerative joint diseases, and recently it has also shown
anti-tumor activity, with an important role in the inhibition of
human-colon adenocarcinoma cell-proliferation. \cite{b.zhang} The
molecule of IMC has been characterized by different spectroscopic
techniques, and it has  been observed that, in the crystalline
state, it exhibits three polymorphic forms (called $\alpha$,
$\beta$ and $\gamma$, respectively), depending on the sample
history.\cite{b.andronis} Due to its chemical structure, IMC is
poorly soluble in water, resulting in high local drug
concentrations responsible for ulcerations in the tissues, and
this reduces its therapeutic applications. A strategy to affect
the solubility and chemical stability of IMC in water, which is
actually employed in commercial drugs, consists in the preparation
of inclusion complexes with CD.

Although some models were proposed for the arrangement of the
guest molecule into the cavity, \cite{b.myles, b.fronza}
unequivocal conclusions were not reached. Moreover, although the
vibrational spectra of the various forms of free IMC have been
widely studied in the literature by FT-IR and Raman spectroscopy,
\cite{b.taylor} no Raman data have been reported on its inclusion
complex with CD.

In the present work we discuss the results of Raman scattering
experiments and numeric simulations on the IMC-CD inclusion
complexes in solid state, which provide new insight into the
structure of the complexes, and into the effect of the inclusion
process on the guest. Additional information is obtained by the
comparison of these results with those relative to (i)
5-metoxyindol-3-acetic acid, (ii) the inclusion complexes formed
by CD with 4-clorobenzoic acid and (iii) with the sodium salt of
IMC (NaIMC). Additional experiments of electrospray-ionization
mass spectrometry and NMR were also performed.
\section{Experimental and computational details}
All chemical compounds were purchased from Sigma-Aldrich and used
without further purification. Raman spectra were recorded using a
Jobin-Yvon HR800 micro-Raman confocal system with a 100$\times$
objective, equipped with a liquid-nitrogen-cooled CCD detector.
Exciting radiation at $632.8$ nm ($20$ mW) was provided by a He-Ne
laser. The Raman spectra were generally recorded using dried
samples. In some cases, when the amount of available complex was
small, in order to improve the Raman signal, Surface Enhanced
Raman Scattering (SERS) was used. In these cases, drops of liquid
sample were repeatedly deposited on a suitable metallic substrate
(Ti-6Al-4V alloy) and dried by evaporation, thereby increasing the
complex concentration. \cite{b.sers1, b.sers2, b.sers3}
Mass spectrometry experiments were performed on a Bruker
Esquire-LC$^{TM}$ ion trap mass spectrometer equipped with an
atmospheric-pressure Electrospray ionization (ESI) source, in the
voltage range 99-180 V; the total ion current was acquired with a
scan range \emph{m/z} 100-2000. The sample was directly infused as
a $1:1$ MeOH/H$_2$O solution at a flow rate of 2 $\mu$l/min into
the ESI source in either positive- or negative-ion mode.
$^{1}$HNMR spectra were taken at room temperature with an Avance
400 Bruker spectrometer; $^{1}$H was recorded at 400 MHz, with
$\delta$ values in ppm in D$_{2}$O relative to the HDO residual
signal ($\delta_H$ 4.70 ppm) and J values in Hz.

All the \emph{ab initio} quantum chemical calculations were
performed with the GAUSSIAN 03 program suite, \cite{b.gaussian}
utilizing unrestricted DFT \cite{b.dft} and the non-local B3LYP
functional hybrid method was employed,\cite{b.dft1} as well as
unrestricted Hartree-Fock (UHF). The standard 6-31G(d) basis set
\cite{b.base} was used for the geometry optimization and frequency
analysis of free indomethacin base pairs. Vibrational frequencies
were computed at the same level of theory, as well as IR and Raman
spectra. In addition, we have also obtained the amplitude of the
displacements of the atoms for the different normal modes. When
the vibrational frequencies have been computed, the scaling factor
of 0.96 \cite{b.wong} was taking into account; this has to be
introduced to correct for anharmonicity, as well as for the
systematic errors inherently present in the calculated harmonic
contributions \cite{b.pople}.

Molecular dynamics simulations were performed by utilizing the
Amber package. \cite{b.amber} In particular, the molecular
dynamics runs have been performed via the module {\em sander},
while the energy minimization and the search for eigenvalues and
eigenvectors of the dynamical matrix have been obtained via the
module {\em nmode}. The force field employed is {\em GAFF}
(Generalized Amber force field~\cite{b.gaff}). In some runs, the
solvent effects have been incorporated explicitly, by including
TIP3 water molecules.~\cite{b.TIP3} The starting structures of IMC
and $\beta$CD have been obtained from the Protein Data
Bank.~\cite{b.PDB} The coordinates for hydrogen atoms were
generated by means of the {\em WWW PRODRG} server.~\cite{b.PRODRG}
Partial charges were computed by means of the module {\em
antechamber} of the Amber package, utilizing the {\em AM1-BCC}
model.~\cite{b.akalian}\\

The IMC-CD complex was prepared according to the procedure adapted
from the literature. \cite{b.lu} $\beta$CD (50.3 mg, 44.4
$\mu$mol) was dissolved in water (4.2 ml) obtaining a 10.6$~\mu$M
solution, which was added to an equimolar amount of dry IMC (15.9
mg, 44.4 $\mu$mol). The resulting dispersion was stirred at
60$^{\circ}$C for 2 hours and cooled at room temperature. After 3
hours, the presence of the complex was indicated by the
precipitation of a white solid, which was collected by drawing the
liquid phase with a syringe, dried in a vacuum chamber (over
P$_{2}$O$_{5}$ as dehydratant), and finally powdered. The
complexes between hydroxypropyl $\beta$CD (HP$\beta$CD) and IMC,
and between $\beta$CD and 4-clorobenzoic acid or sodium salt of
IMC, were obtained under the same procedure.

By $^{1}$HNMR analysis in D$_{2}$O on the complexes of both IMC
and NaIMC with $\beta$CD or HP$\beta$CD, we have obtained the same
$^{1}$H chemical shifts previously reported by Myles
\cite{b.myles} and Fronza.\cite{b.fronza}
\section{Experimental results }
\begin{figure}
\centering
\includegraphics[width=.4\textwidth]{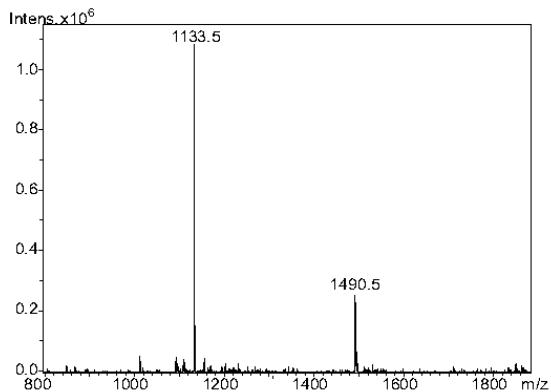}
\caption{ESI(-)-MS spectrum for the IMC-$\beta$CD complex.}
\label{f.2}
\end{figure}

Before the Raman scattering measurements, preliminary experiments
were performed on the complexes using mass spectrometry. In fact,
when applied to the analysis of CD complexes, the negative ion
ESI-MS technique provides information on the stoichiometry of the
complex in gas-phase. In the case of the IMC-$\beta$CD complex, as
shown in Fig.~\ref{f.2}, the cluster signals of the
pseudo-molecular ion [M$_{complex}$-H]$^{-}$ of the 1:1 complex is
detected at \emph{m/z}=1490, and the signal of the ion
[M$_{CD}$-H]$^{-}$ of free $\beta$CD at \emph{m/z}=1133. No
signals arising from $1:2$ complex were observed. Tandem MS/MS
experiments, carried out in the ion trap through isolation of the
cluster of the $1:1$ complex at $1490$ \emph{m/z}, and CID
fragmentation analysis, yielded the daughter ion at \emph{m/z}
$1133$ by neutral loss of IMC. Similar behavior was observed for
IMC-HP$\beta$CD and IMC sodium salt-$\beta$CD complexes,
supporting 1:1 stoichiometry of the complexes in these cases as
well.

\begin{figure}[h!]
\centering
\includegraphics[width=.45\textwidth]{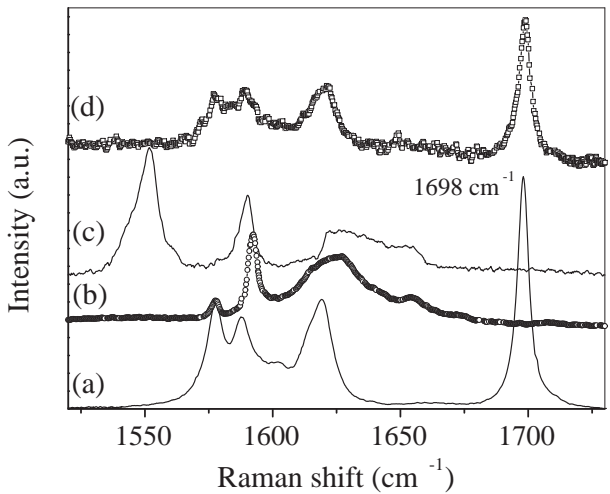}
\caption{Top: Raman spectra of indomethacin (a), 4-chlorobenzoic
acid (b), 5-metoxyindol-3-acetic acid (c), and 1:10 physical
mixture of IMC and $\beta$CD (d). Throughout the paper, different
spectra are vertically shifted for the sake of clarity.}
 \label{f.3}
\end{figure}

\begin{figure}[t]
 \centering
\includegraphics[width=.45\textwidth]{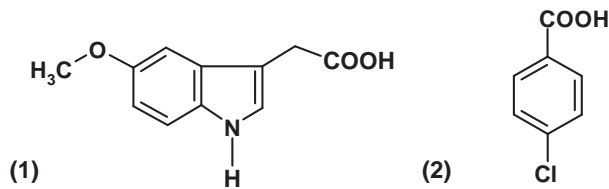}
\caption{ Chemical structure of 5-metoxyindol-3-acetic acid
\textbf{(1)}, and 4-chlorobenzoic acid \textbf{(2)}.}
 \label{f.89}
\end{figure}

In Fig.~\ref{f.3} we report the Raman spectra of solid
$\gamma$-IMC (a), 4-chlorobenzoic acid (b), 5-metoxyindol-3-acetic
acid (c), and the 1:10 physical mixture of IMC and $\beta$CD (d),
in the energy range 1500-1750 cm$^{- 1}$. The two latter molecules
are of interest in the present study because their chemical
structures are very similar to subunits of IMC, as shown in
Fig.~\ref{f.89}.

\begin{figure}[h!]
\centering
\includegraphics[width=.45\textwidth]{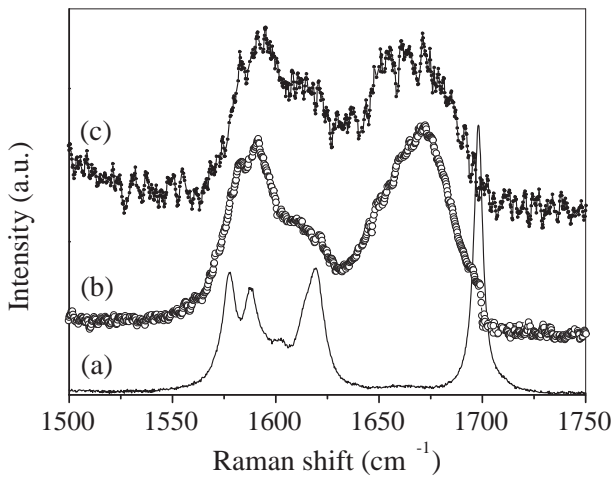}
\caption{Raman spectra of free IMC (a), IMC-HP$\beta$CD complex
(b), IMC-$\beta$CD complex (c).}
 \label{f.5}
\end{figure}

\begin{figure}[h!]
\centering
\includegraphics[width=.45\textwidth]{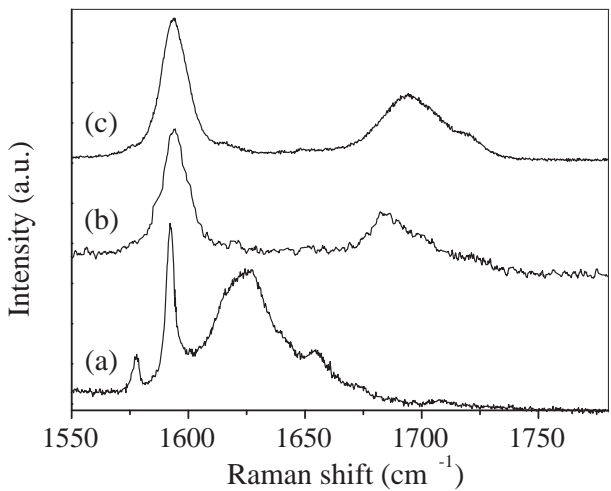}
\caption{Raman spectra of free 4-chlorobenzoic acid (a); 1:10
physical mixture of 4-chlorobenzoic acid-$\beta$CD, respectively
(b); 4-chlorobenzoic acid-$\beta$CD inclusion complex (c).}
\label{f.6}
\end{figure}

We focused on the this energy range because it contains no
free-$\beta$CD peaks, and complexation-induced changes on IMC
spectrum can be readily observed. Moreover, this is also the range
where C=O stretching energies are expected, so that combining
experiment and simulations, useful information on the geometry of
the complex (for instance, the position of the involved CO group)
may be obtained from the features of Raman spectrum.
IMC contains both a C=O amide group (atoms N28-C29-O30 in
Fig.~\ref{f.1}) and a C=O carboxylic group (atoms
C17-O18-O19-H20), which are spatially well separated in the
molecule. The comparison among the Raman spectra of Fig.~\ref{f.3}
seems to indicate that, actually, the peak of IMC at 1698
cm$^{-1}$ corresponds to the amide C=O stretch. In fact, this peak
does not appear in the spectra of compounds \textbf{(1)} and
\textbf{(2)} of Fig.~\ref{f.89}), which do not contain the amide
group; it is also be noted, that this assignment is in agreement
with Taylor et al.. \cite{b.taylor} Moreover, in the following we
shall see that {\em ab initio} quantum chemical calculations and
molecular dynamics computations confirm this conclusion. Thus we
may focus on the peak at 1698 cm$^{-1}$ (from now on the
corresponding frequency will be labelled $\omega_{CON}$, while the
frequency of the carboxylic group will be labelled
$\omega_{COOH}$), and on its changes due to the inclusion process,
which are shown in Fig.~\ref{f.5}. By comparing the spectrum of
free IMC (a) to those of the complexes formed by HP$\beta$CD (b)
and $\beta$CD (c), we note a marked broadening of the peak at
$\omega_{CON}$, as well as a shift from $\approx$ 1700 cm$^{-1}$
to $\approx 1670 {\rm cm}^{-1}$. Such a large shift for
$\omega_{CON}$ ($\approx 30$ cm$^{-1}$) is rather unusual in the
study of the complexation processes where, generally, smaller
 variations are observed. \cite{b.iliescu,b.choi,b.redenti}
 From the above results one might infer that it is the amide C=O group of
IMC (and the neighboring atoms) to be mainly affected by the
inclusion process. However, in order to understand the precise way
in which such atoms are influenced by complexation, other possible
effects should be discussed. Generally speaking, changes of
vibrational frequencies might be due to the presence of the
uncomplexed guest in dimeric form. This is, for example, what
happens to the 4-chlorobenzoic acid which, in the crystalline
state, may exist also in dimeric form due to intermolecular
hydrogen-bonded interactions, and whose Raman spectrum is shown in
Fig.~\ref{f.3} (b): here, the broad peak at $1628$ cm$^{-1}$ has
been assigned to the carboxylic C=O stretch of the dimer,
\cite{b.lee} and thus corresponds to $\omega_{COOH}$. In
Fig.~\ref{f.6} we report the spectra of free 4-chlorobenzoic acid
(a), of the $1:10$ physical mixture of 4-chlorobenzoic
acid-$\beta$CD, respectively (b), and of the $\beta$CD-complexed
4-chlorobenzoic acid (c). In case (b), only
 4-chlorobenzoic acid monomers are expected: as a matter of fact,
  the breakdown of
hydrogen bonding patterns of 4-chlorobenzoic dimers is observed in
the shift of the C=O stretch from 1628 cm$^{-1}$
(Fig.~\ref{f.6}(a)) to higher energy ($\approx 1684$ cm$^{-1}$)
(Fig.~\ref{f.6}(b)). Moreover, the comparison between
Fig.~\ref{f.6} (b) and (c) shows that the complexation process
further shifts the peak corresponding to C=O stretch by $\sim 10$
cm$^{-1}$. Note that this high-energy shift for the
4-chlorobenzoic acid is at odds with the IMC case, where a
decrease of $\omega_{CON}$ was observed upon complexation.

The case of the 4-chlorobenzoic acid shows that complexation may
affect the Raman peaks in at least two ways: (i) by breaking the
intermolecular bonds, (ii) by modifying the strength of the
interatomic bonds (C=O in this case). In order to check the
possibility that the former effect may affect the complexed form
of IMC, we have measured the Raman spectrum of a $1:10$ physical
mixture of IMC-$\beta$CD (Fig.~\ref{f.3} (d)).
  In the energy range between $1500$ and $1750$ cm$^{-1}$, this
  spectrum exhibits the same characteristic peaks as in Fig.~\ref{f.3} (a),
indicating that the shift of the amide C=O stretch, observed in
Fig.~\ref{f.5} (a) and (b), is actually related to complexation
and not to the cleavage of hydrogen bonding patterns of IMC
dimers.

\begin{figure}
\centering
\includegraphics[width=.45\textwidth]{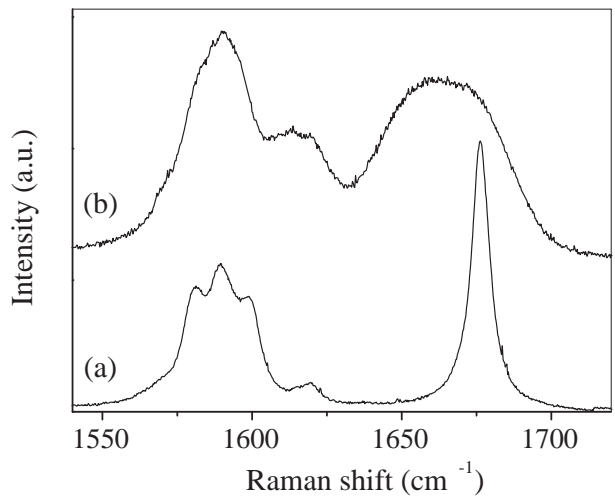}
\caption{Raman spectra of  free IMC sodium salt (a), and IMC
sodium salt-$\beta$CD complex (b).}
 \label{f.7}
\end{figure}

Similar results have been obtained on the inclusion complex formed
by $\beta$CD with IMC sodium salt (Fig.~\ref{f.7}): also in this
case, we clearly observe a broadening and shift to lower energy of
the peak at 1675 cm$^{-1}$, which is generally assigned to the
amide C=O stretching, when the guest is included in $\beta$CD. All
these results confirm the trend already observed in the
vibrational spectra of IMC complexes, and further support the
hypothesis that the amide C=O of IMC is directly involved in the
process of complexation with $\beta$CD.
Finally, the vibrational properties of IMC-$\beta$CD inclusion
complexes have been also investigated in the low-frequency region
(10-100 cm$^{-1}$), using a Jobin Yvon U1000 Raman spectrometer
with Ar-laser excitation at 514.5 nm; preliminary results seem to
evidence, in this spectral region as well, the presence of
significant differences, probably due to extended vibrations of
the system. \cite{b.philmag}

Actually, there is also a third phenomenon that might induce
changes in the experimental spectra, namely the possible change of
the solid state of IMC due to the complexation. In general, below
its glass transition temperature IMC crystallizes in the stable
$\gamma$-form, whereas at higher temperatures the formation of the
metastable $\alpha$-form is predominant.~\cite{b.andronis} The
polymorphic $\alpha$ and $\gamma$ forms differ not only in the
packing density of the crystals, but also in the unit cell
arrangements, \cite{b.chen} leading to a deep difference in their
Raman spectra. \cite{b.taylor} Thus, one could argue that the
large structural change undergone by the solid state due to
complexation, could be related to the observed shift of
$\omega_{CON}$. In order to investigate this possibility, we
performed numeric simulations describing the complexation process
between a single molecule of IMC and a single molecule of
${\beta}CD$. We shall see that they reproduce sufficiently well
the experimental findings and that structural changes do not need
to be taken into account.
\section{Numeric simulations}
\subsection{\emph{Ab initio} quantum chemical computation}
\begin{figure}[h!]
\centering
\includegraphics[width=.45\textwidth]{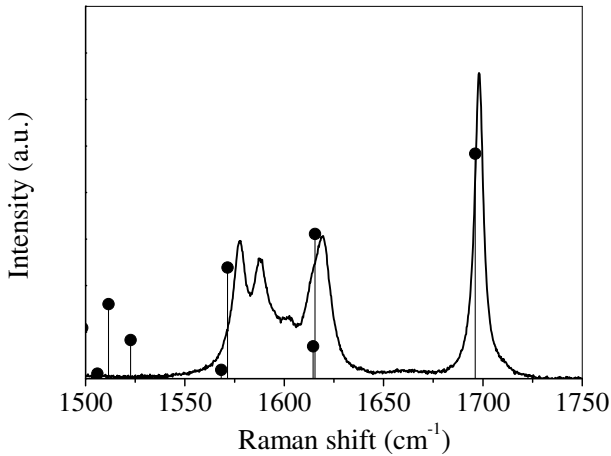}
\caption{Full line: experimental Raman spectrum of free IMC; Full
circles: Raman intensities as computed by UHF.}
 \label{f.uhf}
\end{figure}

Focusing only on the energy range described in the experimental
section, the main result from the {\em ab initio} computation is
that, in the region $1500-1750$ cm$^{-1}$ (a complete assignment
of frequencies for free IMC has been attempted in Ref.
\cite{b.vibrational}), the eigenvector with the largest Raman
signal involves the largest displacement over the atoms C29-O30,
i.e. it corresponds to the amide C=O stretching. The value for
$\omega_{CON}$ found from the UHF {\em ab initio} computation
(1696 cm$^{-1}$ ) is rather close to the experimental one, as
shown in Fig.~\ref{f.uhf}. Furthermore, in the comparison between
experimental and computed Raman intensities, the signals falling
in the region between about $1600$ and $1630$ cm$^{-1}$ can
probably be assigned to aromatic C=C stretch bond of IMC. The
vibrational analysis of complexes has been performed using
classical simulation (see next section), which allows larger
systems to be simulated with respect to the {\em ab initio}
calculation.
\subsection{Molecular dynamics computation}
The advantage of numeric simulations, with respect to the
experiments, is that the positions of the IMC and CD atoms can be
located exactly. On the other hand, numeric simulations of the
complex are rather CPU-time demanding, especially when the effects
of the solvent are taken into account. The length of our runs
ranged from 1 ms, for a molecule of IMC in vacuum, to only a few
ns when thousands of molecules of water were added. To answer in a
reliable way the question about the geometry of the inclusion
process, one should perform long enough runs of many IMC and
$\beta$CD molecules, in the presence of a solvent, in order to
detect many spontaneous inclusion events, and then perform a
statistical study. As we shall see, because of the long CPU-time
needed, this goal has been only marginally achieved, since only
one of such events has been observed. On the other hand, it has
been possible to gain further physical insight by supporting this
result with somewhat less ambitious simulations.
\begin{figure}
\centering
\includegraphics[width=.35\textwidth]{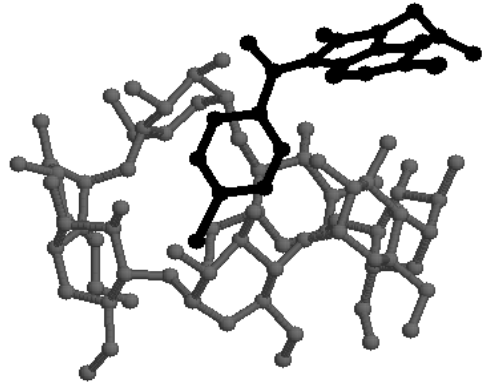}
\caption{Starting configuration of the IMC-$\beta$CD complex
utilized in cases (3) and (4) (see text). In the former case,
molecules of water have been added.}
 \label{f.8}
\end{figure}

In order to study the inclusion of IMC in $\beta$CD we have
simulated the following systems: (1) single IMC molecule in
vacuum, (2) single IMC molecule in a box of 1114 molecules of
water, (3) single IMC-$\beta$CD inclusion complex in vacuum, (4)
single inclusion complex in a box of 633 molecules of water and
(5) 9 IMC molecules plus 9 $\beta$CD molecules in a box of 1400
molecules of water. In systems (3) and (4) the inclusion complex
was prepared by docking the Cl atom of IMC into the cavity of
$\beta$CD, and minimizing the energy of the resulting
configuration (see Fig.~\ref{f.8}). During the molecular dynamics
runs, we recorded the distance $d$ between the Cl atom of IMC and
the center of mass of $\beta$CD, in order to determine the
formation of an inclusion complex. More precisely, a complex is
defined by the fact that $d \sim 0$ (or, at least, much lesser
than the typical distance between free IMC and $\beta$CD).
Moreover, starting from the thermodynamic equilibrium
configuration, we sought for the minima of the energy potential.
In the minimum configurations we computed the eigenvalues $\omega$
and the corresponding eigenvectors $\vec e_i(\omega)$ of the
dynamical matrix, in order to determine the vibrational properties
of the different systems. From such quantities we have obtained
the density of vibrational states $g(\omega)$ and the quantity
$Y_i \equiv \vec e_i \cdot \vec e_i$, which measures the amplitude
of the vibration of the $i$-th atom due to the mode of frequency
$\omega$. The thermodynamic parameters chosen for the simulations
are room temperature and room pressure, with the coupling to the
external bath introduced via a weak coupling
scheme.~\cite{b.berendsen}
\begin{figure}[h!]
\centering
\includegraphics[width=.35\textwidth,angle=270]{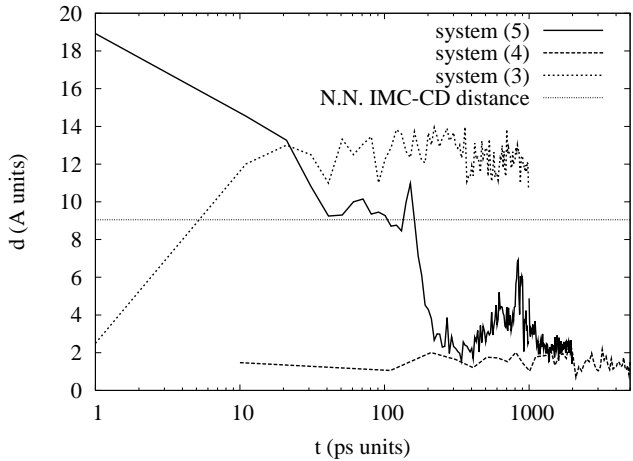}
\caption{Distance $d$ between the Cl atom  of IMC, and the center
of mass of $\beta$CD, as a function of time for different
simulated
  systems (see text).}
 \label{f.9}
\end{figure}
In Fig.~(\ref{f.9}) we show the behavior of the distance $d$
during different molecular dynamics runs. The solid line
represents the distance between the Cl atom of IMC and the center
of mass of $\beta$CD, for a given pair chosen among the 9 IMC plus
9 $\beta$CD molecules of system (5). After 1 ns, $d$ reaches a
value of approximately 2 $\AA$, which we interpret as the
signature of an inclusion event, with the Cl located quite close
to the center of mass of $\beta$CD. In our runs, we did not
observe other inclusion processes, thus we are not able to perform
a detailed study of the phenomenon (characteristic times, etc...).
It is interesting to note that $d$ is much smaller than the
position of the first peak of the pair correlation function
between {\em free} IMC and CD molecules (horizontal line).
Furthermore, the value of $d$ reached in the spontaneous inclusion
process found in system (5) is very similar to the stationary
value of the distance in the run of system (4) (dashed line),
where the complex was initially created and put in a box of water.
Both results support our claim that in system (5) (which is the
most similar to an experimental sample) we actually observed a
single complexation process. Interestingly, the complex is not
stable without water: in fact, we see that although system (3)
(dotted line) starts from the same initial configuration as system
(4), the absence of water leads to an immediate destruction of the
complex.
\begin{figure}[h!]
\centering
\includegraphics[width=.7\columnwidth,angle=270]{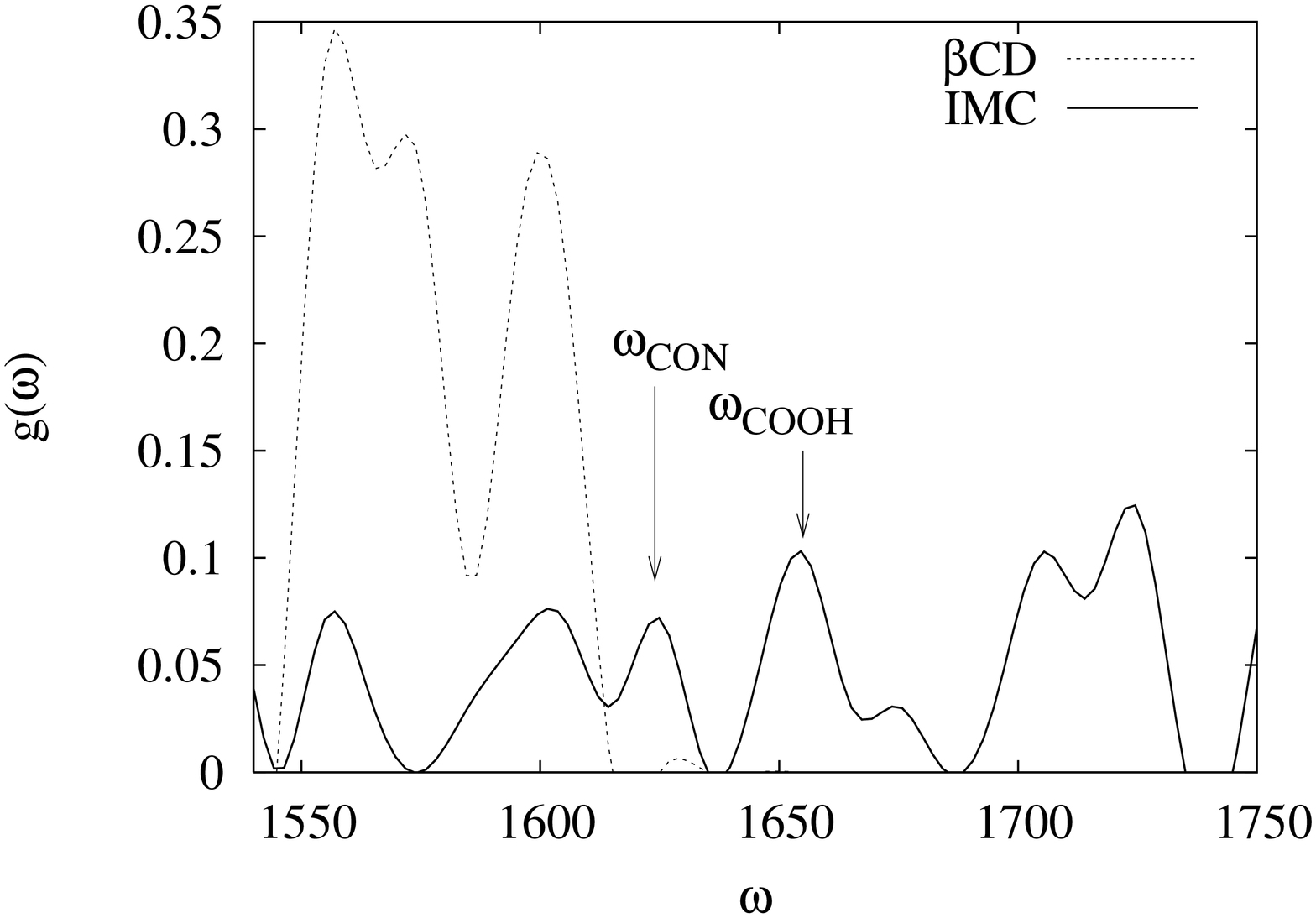}
\includegraphics[width=.7\columnwidth,angle=270]{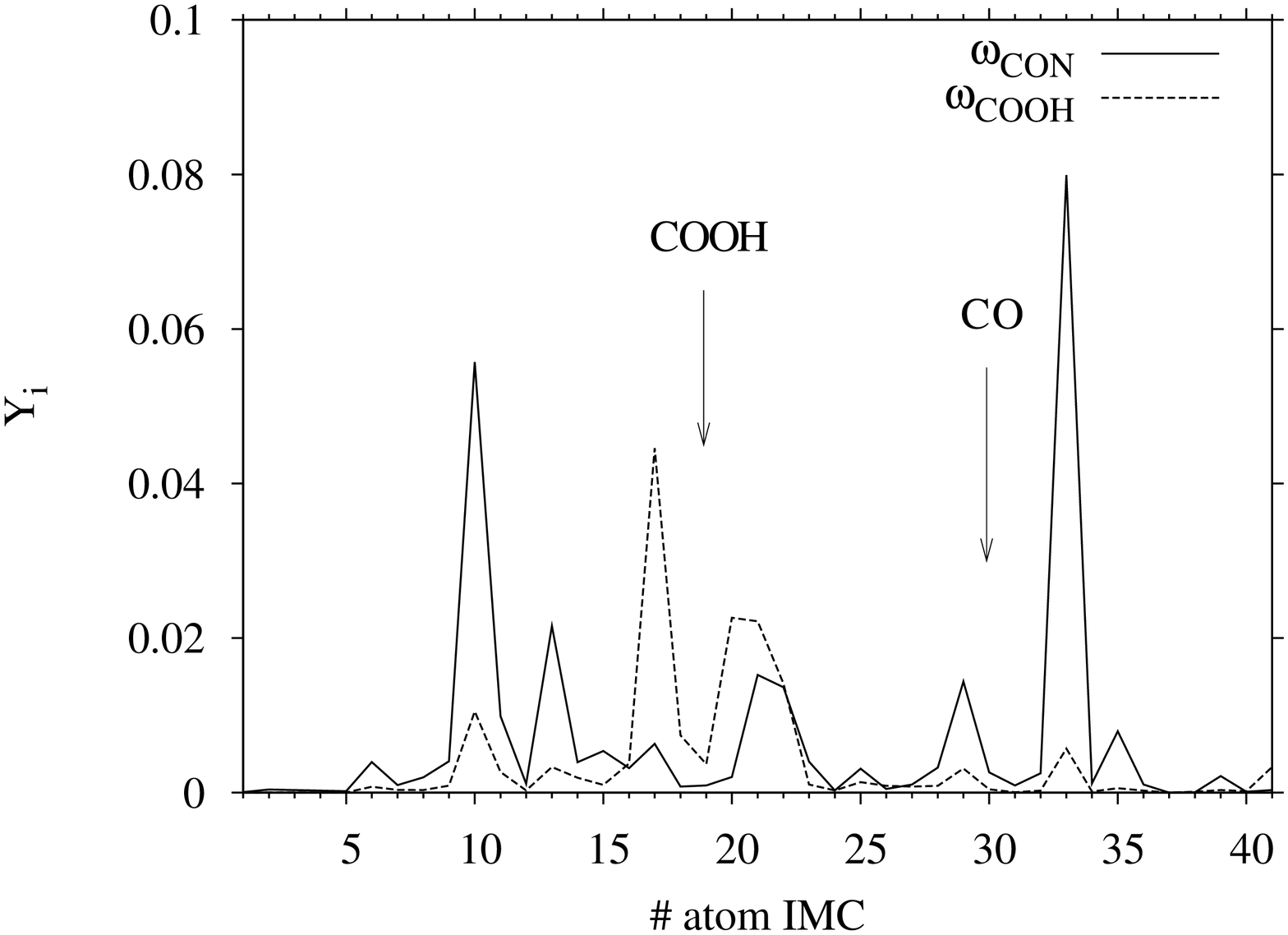}
\caption{Top: density of vibrational states $g(\omega)$ for the
simulated free IMC (solid line) and $\beta$CD (dashed line).
Bottom: projection of the
  eigenvectors corresponding to $\omega_{CON}$ and $\omega_{COOH}$ over
  the atoms of IMC, labelled as in Fig. 1. Note that in both cases
  there are also
  large contributions from some hydrogen atom. }
 \label{f.11}
\end{figure}

Unlike the {\em ab initio} quantum chemical computations discussed
previously, from classical molecular dynamics (MD) simulations it
is not possible to obtain in a straightforward way the Raman
spectra; in fact, the Raman signal $I(\omega)$ is $\propto
g(\omega) \, C(\omega)$, where $C(\omega)$ is the so-called
coupling function which, in classical computations, is unknown
unless the mechanism of polarizability modulation is specified.
\cite{b.anderson} On the other hand, with MD simulations one can
obtain the eigenfrequencies and the eigenvectors of the inclusion
complex.

The interpretation of the density of vibrational states
$g(\omega)$ proceeds through different steps. In Fig.~\ref{f.11}
we identify the frequencies $\omega_{CON}$= 1628 cm$^{-1}$ (Raman
active according to the ab initio calculation, see Section IV A)
and $\omega_{COOH}$= 1649 cm$^{-1}$ (Raman inactive), as the ones
corresponding to the modes which are most localized on the atoms
of the amide and of the carboxylic C=O groups, respectively. There
is a slight difference between the values of these two
characteristic frequencies and the ones computed with the {\em ab
initio} study, and between $\omega_{CON}$ as computed with the MD
approach and the experimental value. These discrepancies might be
ascribed to the force field utilized.

\begin{figure}[h!]
\centering
\includegraphics[width=.7\columnwidth,angle=270]{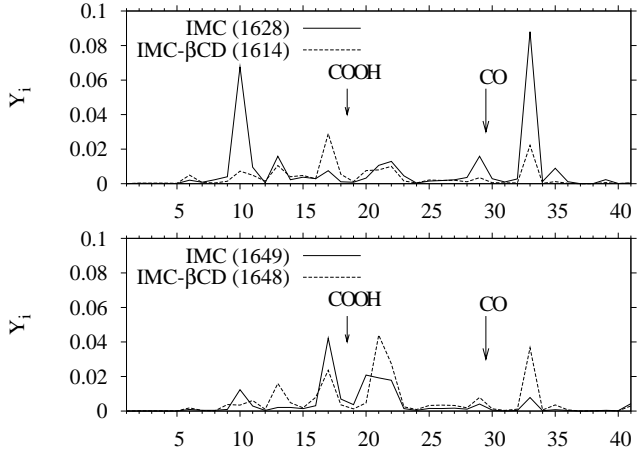}
\caption{Top: projection of the eigenvector
  corresponding to $\omega_{CON}$ over the atoms of IMC (labelled as in Fig. 1),
  in free IMC (solid line, system (1))
  and IMC-$\beta$CD complex (dotted line, system (4)).
  Bottom: projection of the eigenvector
  corresponding to $\omega_{COOH}$ in free IMC (solid line, system (1))
  and IMC-$\beta$CD complex (dotted line, system (4)).}
 \label{autocomplesso}
\end{figure}
In Fig.~\ref{autocomplesso} is shown the effect of complexation on
the eigenvectors corresponding to $\omega_{CON}$ and
$\omega_{COOH}$. In fact, in the complexed system (4), we have
looked for the frequencies whose corresponding eigenvectors were
the most similar to the ones of free IMC shown in Fig.~\ref{f.11}.
While $\omega_{COOH}$ is not affected by the inclusion process,
one can see that $\omega_{CON}$, corresponding to the largest
experimental Raman signal, shifts to lower frequencies by $\sim
14$ cm$^{-1}$. Thus, from a qualitative point of view, we find
results in agreement with the experimental findings described in
the above section. Quantitatively, one has to note that this shift
is about one half of the experimental one.
%
\begin{figure}[h!]
\centering
\includegraphics[width=.7\columnwidth,angle=270]{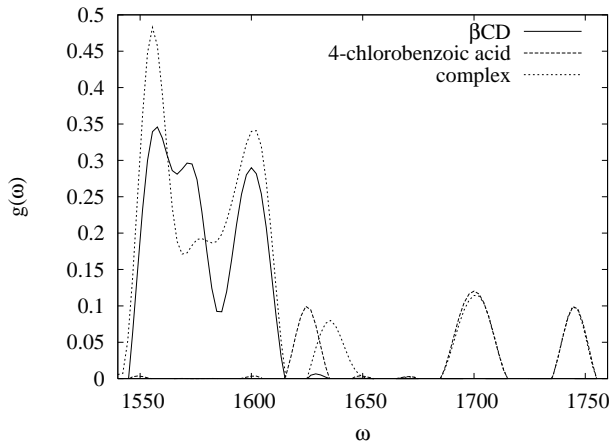}
\caption{Density of vibrational states $g(\omega)$ for free
  $\beta$CD (solid line), for 4-chlorobenzoic acid (dashed line),
  and for $4$-chlorobenzoic-$\beta$CD-complex (dotted line).}
\label{acido}
\end{figure}
It seems reasonable to ascribe this difference to the limitations
of the GAFF force field in describing a molecule of the complexity
of IMC, rather than to the fact that in these simulations the
structural properties of the solid state (and their possible
modifications due to the complexation) are not taken into account.
As a matter of fact, in Fig.~\ref{acido} we show that for a much
simpler molecule like the 4-chlorobenzoic acid (for which the GAFF
force field is expected to be more reliable), the MD simulations
show a shift of $\omega_{COOH}$ to higher energy of $\sim 10$
cm$^{-1}$, which is in good agreement with the one found in
experiments in the complexation of the monomeric form of
4-chlorobenzoic acid.
\section{Conclusions}
In the present paper we have reported the results of a
multi-technique study of the complexes formed by CD with the
molecule of IMC. In particular, we have performed Raman scattering
experiments and we have compared the experimental findings with
the outcome of {\em ab initio} quantum chemical calculations and
of molecular dynamics simulations. Focusing on the energy range
$1500-1700$ cm$^{-1}$, where Raman spectra show the largest
modifications after the complexation process, we conclude that the
geometry of the IMC-$\beta$CD complex is the one shown in
Fig.~\ref{f.8}, characterized by the 4-chlorobenzoyl unit inserted
into the cavity of $\beta$CD through its larger rim, as early
proposed by Fronza et al..\cite{b.fronza} We have also computed
the density of vibrational states of free IMC, of $\beta$CD, and
of their inclusion complex, and compared it to the corresponding
experimental Raman spectra. Though it is difficult to reach
complete quantitative agreement between experiment and
simulations, the latter explain the shift to lower frequencies of
the most intense Raman peak, whose eigenvector is localized on the
atoms of the amide C=O group; this shift is due only to the
inclusion process, whereas dimerization processes or structural
changes in the solid state seem not to be involved. This
multi-technique, experimental and numerical approach, appears very
promising for the study of other inclusion complexes as well.
\section{Acknowledgments}
We would like to thank G.~Mariotto for making the micro-Raman
apparatus available, A.~Sterni for technical assistance in the MS
analysis and F.~Pederiva for useful discussions.

\end{document}